\documentclass[prl,aps,twocolumn,superscriptaddress,showpacs]{revtex4}
\usepackage{amsmath,amssymb,graphicx}
\usepackage{pxfonts}
\usepackage{graphicx}
\begin{document}

\title{Collection of indirect excitons in a diamond-shaped electrostatic trap}

\author{A.~A. High}
\author{A.~K. Thomas}
\author{G. Grosso}
\author{M. Remeika}
\author{A.~T. Hammack}
\author{A.~D. Meyertholen}
\author{M.~M. Fogler}
\author{L.~V. Butov}
\affiliation{Department of Physics, University of California at San
Diego, La Jolla, CA 92093-0319}

\author{M. Hanson}
\author{A.~C. Gossard}
\affiliation{Materials Department, University of California at Santa
Barbara, Santa Barbara, California 93106-5050}

\begin{abstract}

We report on the principle and realization of a new trap for excitons -- the diamond electrostatic trap -- which uses a single electrode to create a confining potential for excitons. We also create elevated diamond traps which permit evaporative cooling of the exciton gas. We observe collection of excitons towards the trap center with increasing exciton density. This effect is due to screening of disorder in the trap by the excitons. As a result, the diamond trap behaves as a smooth parabolic potential which realizes a cold and dense exciton gas at the trap center.

\end{abstract}


\date{\today}

\maketitle

Potential traps are an effective tool for studying the physics of cold atoms. They made possible the realization of cold and dense atomic gases and, eventually, the achievement of atomic BEC. Furthermore, they allow precise control of atomic gases by allowing in-situ control of trap shape and depth \cite{Cornell02,Ketterle02}.

Traps can become an effective tool for studying the physics of cold excitons -- cold bosons in condensed matter materials. Of particular interest is a trap which can provide a confining potential with the exciton energy gradually reducing towards the trap center. Such a parabolic-like potential could collect excitons from a large area, creating a dense exciton gas. This, in turn, can facilitate the creation of a degenerate exciton gas and eventually exciton BEC in the trap.

Excitons have been studied in a variety of static traps whose profile cannot be changed in-situ, i.e. on a time scale shorter than the exciton lifetime. These traps include strain-induced traps \cite{Trauernicht83,Kash88,Negoita99}, traps created by laser-induced interdiffusion \cite{Brunner92}, and magnetic traps \cite{Christianen98}. Recently, excitons were also studied in traps whose shape and depth can be controlled in-situ -- electrostatic traps \cite{Huber98,Hammack06,Chen06, High09} and laser-induced traps \cite{Hammack06a}.

The realization of a cold and dense exciton gas in a trap requires a long exciton lifetime, which allows the excitons to travel to the trap center and cool to low temperatures before recombination. An indirect exciton in a coupled QW structure (CQW) is composed of an electron and a hole in separate wells (Fig. 1a,b). Lifetimes of indirect excitons are orders of magnitude longer than lifetimes of regular excitons and long enough for the excitons to travel over tens and hundreds of microns and cool below the temperature of quantum degeneracy.

It is also important to understand the role of disorder in the behavior of excitons in a trap. Disorder is an intrinsic feature of solid state materials. In QW structures, it forms mainly due to QW width and alloy fluctuations. Even in a high-quality QW, the in-plane disorder amplitude is on the order of 1 meV, exceeding the thermal energy at low temperatures. Therefore, localization effects should play a significant role.

\begin{figure}
\includegraphics[width=6.5cm]{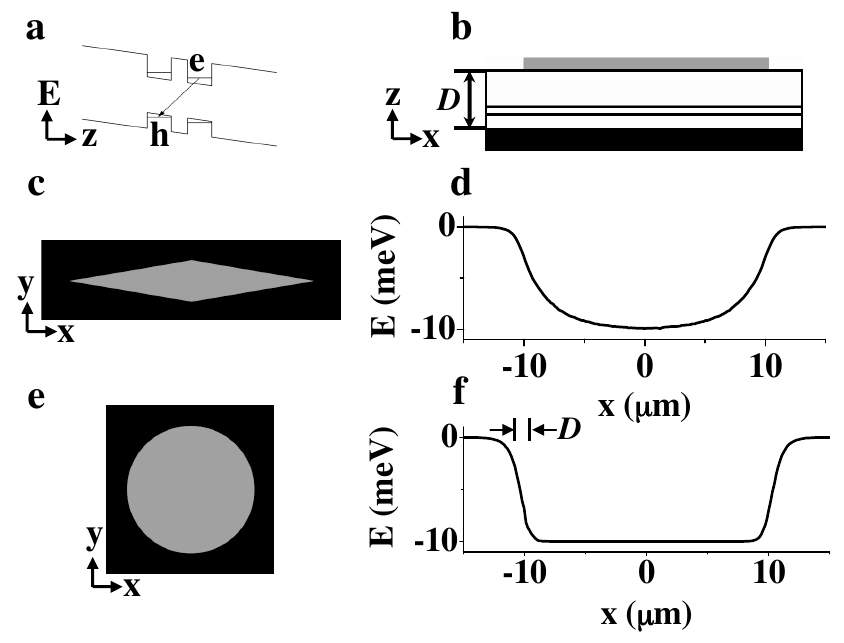}
\caption{(a) Energy band diagram of the CQW; e, electron;
h, hole. (b) Schematic of device structure. (c,d) A diamond-shaped electrode creates a parabolic-like trap. (e,f) A circular electrode creates a box-like trap.  $D=1 \mu$m, the CQW is 100 nm above the bottom electrode, $d=12$ nm, the diamond electrode is $3.5 \times 20 \mu$m, and the applied voltage is 1 V in the calculations.}
\end{figure}

In this paper, we report on the principle and realization of a new electrostatic trap for indirect excitons - the diamond trap (Fig. 1c,d). The diamond trap creates a confining potential for excitons by using a single electrode. At low densities and temperatures, excitons in the trap are localized by the disorder potential. However, with increasing density, the disorder is screened by exciton-exciton interaction, and the excitons become free to collect to the trap center. At high density, the trap behaves as a smooth parabolic potential which realizes a cold and dense exciton gas at the trap center.

The principle of the diamond trap is the following. Indirect excitons have a built-in dipole moment $ed$, where $d$ is close to the distance between the QW centers. An electric field $F_z$ perpendicular to the QW plane results in the exciton energy shift $\delta E = e d F_z$. A laterally modulated electrode voltage $V(x,y)$ creates a lateral potential profile for the indirect excitons $E(x,y) = e d F_z(x,y) \propto V(x,y)$. An appropriately designed voltage pattern permits the creation of a variety of in-plane potential profiles for indirect excitons, including ramps \cite{Hagn95,Gartner06}, lattices \cite{Zimmermann97,Krauss04,Hammack06,Remeika09}, traps \cite{Huber98,Hammack06,Chen06,High09}, and circuit devices \cite{High07,High08}.

Confining potentials can be made by using multi-ring traps \cite{Hammack06}, however such traps require a large number of contacts, which makes the design complex. In contrast, the diamond trap creates a parabolic-like potential using a single electrode. Advantages include simplicity of fabrication and accuracy of control. For instance, switching off the confining potential, like in atomic time-of-flight experiments, or modulating trap depth, like in atomic collective mode experiments, \cite{Cornell02,Ketterle02} can be realized by modulating the electrode voltage.

\begin{figure}
\includegraphics[width=8.5cm]{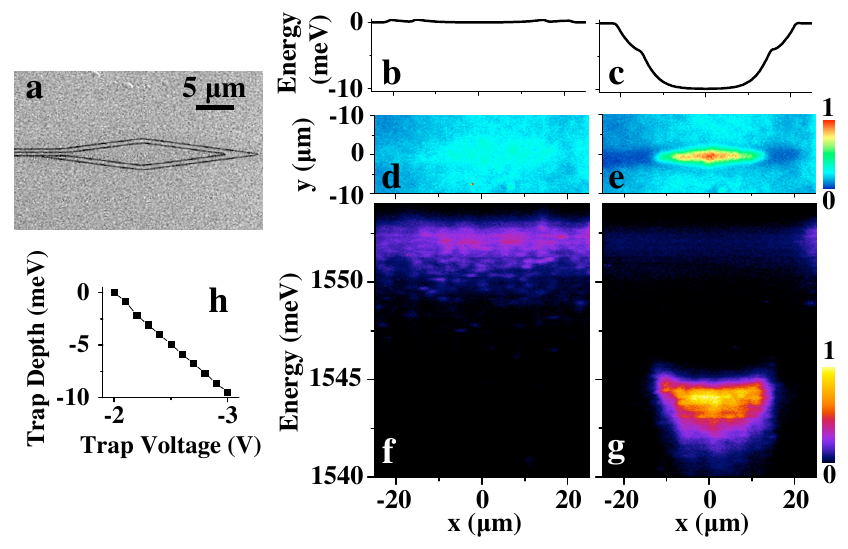}
\caption{(Color online) (a) SEM image of the diamond trap. (b,d,f) Flat potential ($V_d=V_w=V_p=-2$ V). (c,e,g) Normal trap ($V_d =-3, V_w=-2.5, V_p=-2$ V). (b,c) Simulation of exciton energy. (d,e) $x-y$ and (f,g) $x-E$ emission images. (h) Trap depth vs. $V_d$. For all data, $T=1.4$ K, $\lambda_{ex}=633$ nm, $P_{ex} = 28 \mu$W.}
\end{figure}

The diamond trap is formed by a single top electrode of a diamond shape, Fig. 1c. Since a thinner electrode produces a smaller electric field in the $z$ direction (due to the field divergence near the electrode edges), the potential of the diamond trap $e F_z d$ is deepest at the center, where the electrode is widest, and becomes progressively shallower towards the diamond tips. The trap profile calculated numerically from the Poisson equation is presented in Fig. 1d. Its shape is well approximated by
\begin{equation}
U_\text{trap}(x,y)= \frac{-U_{0}}{2}\left[
\tanh{\left(\frac{y+w}{a}\right)} -
\tanh{\left(\frac{y-w}{a}\right)}\right],
\label{eqn:U_trap}
\end{equation}
where $a$ controls the sharpness of the trap and $w=w(x)=L_y(1-|x|/L_x)$ is the width of the diamond pattern. The calculations show that a parabolic-like in-plane potential is formed by a diamond electrode (the potential is also parabolic-like in y-direction). In comparison, the potential created by a circular electrode is box-like (Fig. 1e,f). It varies only within a small length from the electrode edge $\sim D$, where $D$ is the distance between the electrodes, 1 $\mu$m in our samples (Fig. 1b).

The CQW structure was grown by MBE. An $n^+$-GaAs layer with $n_{Si}=10^{18}$ cm$^{-3}$ serves as a homogeneous bottom electrode. The top electrodes on the surface of the structure were fabricated via e-beam lithography by depositing a semitransparent layer of Ti (2 nm) and Pt (8 nm). The device includes a $3.5 \times 30 \mu$m diamond electrode, a 600 nm wide 'wire' electrode which surrounds the diamond, and 'outer plane' electrode (Fig. 2a). Two 8 nm GaAs QWs separated by a 4 nm Al$_{0.33}$Ga$_{0.67}$As barrier were positioned 100 nm above the $n^+$-GaAs layer within an undoped 1 $\mu$m thick Al$_{0.33}$Ga$_{0.67}$As layer. Positioning the CQW closer to the homogeneous electrode suppresses the in-plane electric field \cite{Hammack06}, which otherwise can lead to exciton dissociation. The excitons were photoexcited by a 633 nm HeNe laser or Ti:Sp laser tuned to the direct exciton resonance at 789 nm. The excitation was defocused over the device area. The spectra were measured with resolution $40 \mu$eV. The spatial resolution was $1.5 \mu$m. Potential profiles were calculated using COMSOL Inc. Multiphysics software.

\begin{figure}
\includegraphics[width=8.5cm]{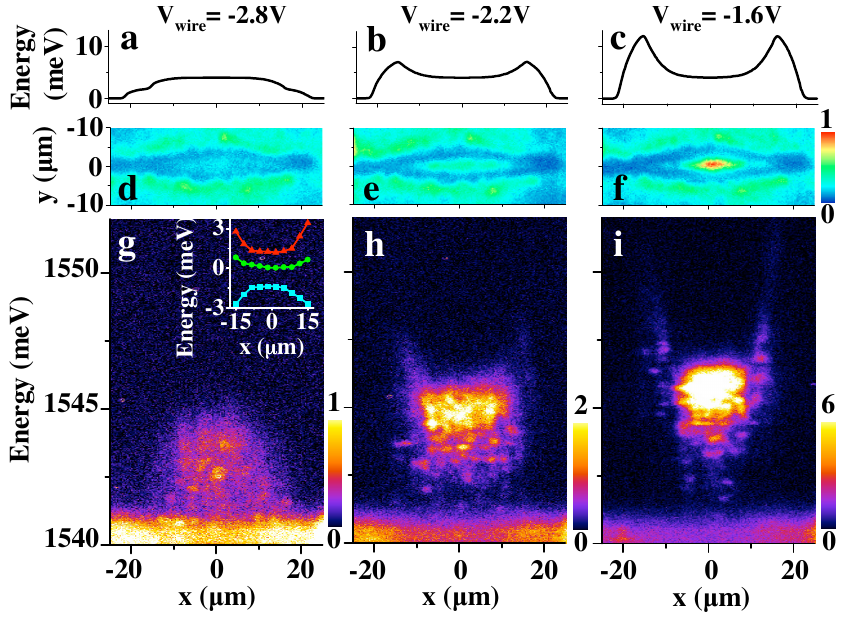}
\caption{(Color online) (a-c) Calculated potential profiles vs. $|V_w|$ showing the transition from a bump to elevated trap. The corresponding $x-y$ (d-f) and $x-E$ (g-i) emission images of the exciton cloud. Inset: the measured trap profile for $|V_w|=2.8$ (squares), 2.0 (circles), and 1.6 V (triangles). For all data, $T=1.4$ K, $\lambda_{ex}=633$ nm, $P_{ex} = 28 \mu$W, $V_d=-2.5$ V, $V_p=-3$ V.}
\end{figure}

Experimental results verify the creation of a parabolic-like trap (Fig. 2). By varying the diamond electrode voltage $V_d$ with respect to the wire voltage $V_w$ and outer plane voltage $V_p$, we go from a flat exciton energy (Fig. 2b,d,f) to a normal trap (Fig. 2c,e,g). The parabolic-like shape of the trap is evident in the $x-E$ emission image (Fig. 2g). The collection of excitons to trap center is seen in both the $x-y$ and $x-E$ images (Fig. 2e,g).

The 'wire' and 'outer plane' electrodes are not necessary for creating a normal trap. However, they are used for creating an elevated trap in which the trap energy is at a higher energy than its surroundings. This allows evaporative cooling of the exciton gas, since higher energy excitons are more likely to escape the trap than lower energy excitons. The evaporative cooling results in an enhancement of the population of lower energy exciton states localized in local minima of the disorder potential in the trap \cite{High09}. The localized exciton states manifest themselves by sharp emission lines located at the state positions (Figs. 3h, 4a).

Figure 3 demonstrates creation of elevated traps. Varying $V_w$ transforms the energy profile from a bump to elevated trap. Increasing the height of the trap walls enhances the exciton collection to the trap center. A higher amount of spatially localized sharp lines in the spectrum shows that excitons are more localized in the elevated trap regime than in the normal trap regime (Fig. 2g, 3h), indicating lower exciton temperature in the former.

The remainder of the paper concerns the elevated trap regime. Figure 4 presents the density dependence. At low density, the excitons are localized in disorder and evenly distributed across the trap (Fig. 4a). As the exciton density is increased, the disorder is screened by exciton-exciton interaction as revealed by the disappearance of sharp emission lines, which correspond to the localized states (Fig. 4a-c). This is quantified by the signal variation due to localization $\sigma$, which drops with increasing density, indicating screening of the disorder (Fig. 4d, the noise contribution to $\sigma$ is negligible).

At high densities, excitons effectively collect to the trap center. This is revealed by a significant reduction of the exciton cloud width (Fig. 4a-c, 5a,b,e). We attribute the observed density-enhanced exciton collection to the screening of trap disorder by exciton-exciton interaction. This interpretation is confirmed by the temperature dependence at low densities (Fig. 5c,d,f): the exciton cloud width decreases with increasing temperature, consistent with exciton delocalization due to thermal activation over the disorder potential.

\begin{figure}
\includegraphics[width=8.5cm]{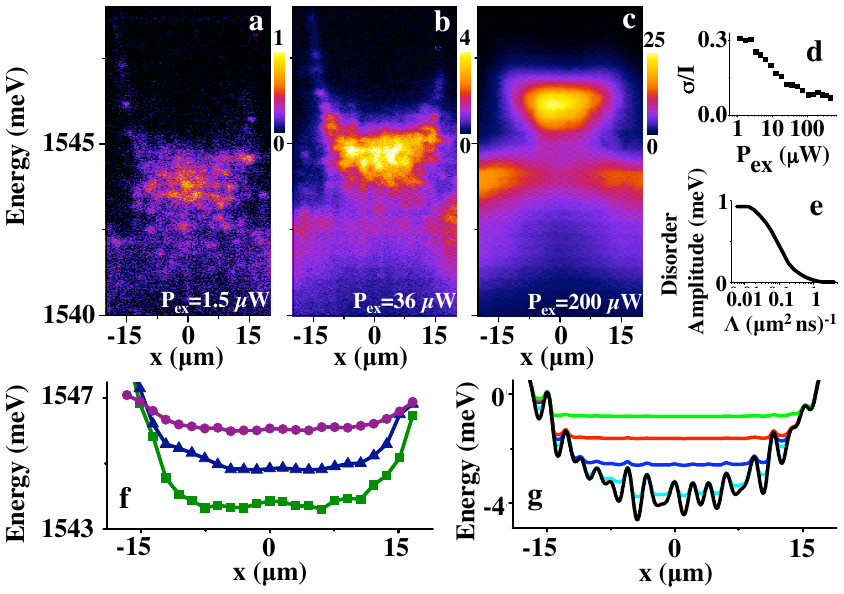}
\caption{(Color online) (a-c) $x-E$ emission images of excitons in an
elevated trap for $P_{ex} = 1.5, 36$, and 200 $\mu$W. (d) Ratio of signal variation to signal intensity. (e) Simulated amplitude of 1 meV disorder with increasing exciton density. (f) Energy profile for $P_{ex} = 1.5$ (squares), 36 (triangles), and $200 \mu$W (circles). (g) Simulated energy profile of indirect excitons in a parabolic-like trap with disorder for exciton creation rate $\Lambda = 0.1$ (cyan), 0.5 (blue), 1.3 (red), and 2 $\mu$m$^{-2}$ ns$^{-1}$ (green). For all data, $T=1.4$ K, $\lambda_{ex}=789$ nm, $V_d=-2.5$, $V_w=-1.8$, and $V_p=-3$ V.}
\end{figure}

This interpretation was examined by theoretical estimates. Similar to Ref. [20], we start with the continuity equation for exciton density
$n(\mathbf{r}, t)$ and current $\mathbf{J}(\mathbf{r}, t)$:
\begin{equation}\label{eqn:diffusion1}
\frac{\partial n}{\partial t} + \nabla \mathbf{J} + R n = \Lambda\,,
\end{equation}
where $R = \tau_r^{-1}$ is the recombination rate and $\Lambda$ is the
generation rate due to photoexcitation. Taking into account the Einstein
relation, we can write the current as follows:
\begin{equation}\label{eqn:current1}
\mathbf{J} = -D \frac{d n}{d\mu_0} \nabla
               \left\{
                        \left[U(\mathbf{r}) + \gamma \nu_1 n \right]
                      + \mu_0(n)
               \right\}\,.
\end{equation}
Here $D$ is the diffusion coefficient. The terms in the square brackets present the potential acting on the excitons. $U(\mathbf{r})$ is the potential due to trap and disorder and $\gamma \nu_1 n$ is the potential created by exciton-exciton repulsion. The latter is treated as local, neglecting $1/r^3$ dipolar tails. $\nu_1$ is the exciton density of states per spin. The dimensionless parameter $\gamma$ presents the interaction strength. The last term is the ``bare'' chemical potential (due to kinetic energy only) given by
\begin{equation}\label{eqn:chemical_potential}
\mu_0(n,T) = T \ln \left(1-e^{-n/n_0}\right)\,,
\quad n_0 = g\nu_1 T\,.
\end{equation}
Here $n_0$ is the density at which exciton gas become quantum degenerate
and $g = 4$ is the spin degeneracy.

We measure density in dimensionless units $\tilde{n} \equiv n/n_0$, which has the physical meaning of the Bose occupation factor. $n_0 \approx 1.5 \times 10^{10}\,\text{cm}^{-2} \times T[\text{K}]$. Then
\begin{gather}
\frac{\partial \tilde{n}}{\partial t} + \nabla \tilde{\mathbf{J}}
+ R \tilde{n} = \frac{\Lambda}{n_0}\,,
\label{eqn:diffusion2}\\
\tilde{\mathbf{J}} = -\frac{D}{T} \left(e^{\tilde{n}}
-1\right) \nabla U - D_\text{eff} \nabla\tilde{n}\,,
\\
D_\text{eff}= D\left[g \gamma \right(e^{\tilde{n}}-1\left)+1\right]\,.
\end{gather}
In the steady state $\partial \tilde{n} / \partial t = 0$, and
Eq.~\eqref{eqn:diffusion2} becomes
\begin{gather}
\nabla \left(-D_\text{eff}\nabla\tilde{n} + \mathbf{J}_\text{con}
       \right)
 + R \tilde{n} = \frac{\Lambda}{n_0}\,,
\label{eqn:diffusion4}\\
\mathbf{J}_\text{con} = -\frac{D}{T} \left(e^{\tilde{n}}
-1\right) \nabla {U}\,.
\end{gather}
This equation can be obtained from formulas of Ref.~\onlinecite{Ivanov_06} with a simplifying approximation of constant $T$ and $R$.

In the model, $U = U_\text{trap} + U_\text{dis}$, with the trap potential
given by Eq. \eqref{eqn:U_trap}. Disorder in our narrow CQW is primarily due to fluctuations of the QW width. Such fluctuations occur on all length scales. The short-range disorder is accounted for by the diffusion coefficient $D$. Potential $U_\text{dis}$ represents the long-range component. It was simulated by three incommensurate lattices $U_\text{dis} = (\delta_{dis}/3) (\sin{\pi\omega_1x} + \cos{\pi\omega_2x} + \sin{\pi\omega_3x}), \omega_1= \sqrt{2}, \omega_2 = \sqrt{3}, \omega_3= \sqrt{5} \mu\,\text{m}^{-1}$. Here $\delta_{dis}$ is an adjustable amplitude. The set of equations (8), (9) was solved numerically for $\tilde{n}(\mathbf{r})$. In the calculations, $\gamma=3$ and $\tau_r=30$ ns following Refs. \cite{Remeika09,Ivanov_06} and the boundary conditions at the diamond perimeter $\tilde{n} = \Lambda / (R n_0) = \text{const}$.  
\begin{figure}
\includegraphics[width=8.6cm]{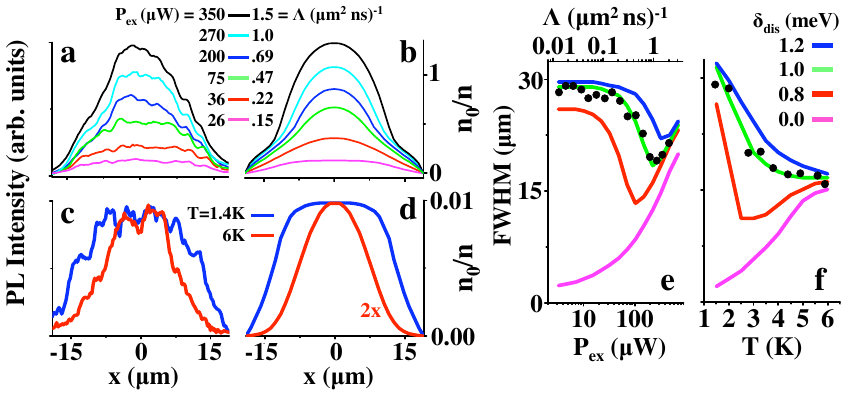}
\caption{(Color online) (a) Experimental emission profile and (b) simulated density profile of excitons in the trap vs. $P_{ex}$ at $T=1.4$ K. $n_0$ is the density of quantum degeneracy. (c) Experimental emission profile and (d) simulated density profile of excitons for $T=1.4$ and 6 K at $P_{ex}=9 \mu$W, $\Lambda = 0.05$. Experimental (points) and simulated (lines) FWHM of the exciton cloud (e) vs. $P_{ex}$ and $\Lambda$ at $T=1.4$ K and (f) vs. $T$ at $P_{ex}=9 \mu$W, $\Lambda$=0.05 for different disorder amplitudes. For all data, $V_d=-2.5$, $V_w=-1.8$, and $V_p=-3$ V. $\lambda_{ex}=789$ nm.}
\end{figure}

The simulations show the screening of disorder by interacting indirect excitons and vanishing of localized states (Fig. 4e,g), in agreement with the experiment (Fig. 4d). Furthermore, the simulations reveal the role of disorder in exciton collection to the trap center. For zero disorder amplitude, the width of exciton cloud monotonically increases with density due to the screening of $U_\text{trap}$. However, for finite disorder, the dependence is nonmonotonic: At low densities, excitons are localized in disorder and broadly distributed over the trap area. As the density increases, excitons begin to screen the disorder $U_\text{dis}$ and collect towards the trap center, which leads to cloud narrowing. At high densities, the disorder is essentially screened, and excitons begin to screen the long range trap potential $U_\text{trap}$, which leads to cloud broadening (Fig. 5e). The screening of the potential is also demonstrated by the energy profiles (Fig. 4f,g).

Increasing the temperature leads to a monotonic increase of the cloud width for zero disorder due to the thermal occupation of higher energy states in the trap. However, for finite disorder, increasing the temperature first leads to the narrowing of the exciton cloud due to thermal delocalization of excitons and collection to the trap center and then to the cloud widening when delocalized excitons spread over higher energy states in the trap (Fig. 5f). The comparison to the experimental data (Fig. 5e,f) leads to an estimated long-range disorder amplitude $\delta_{dis} \sim 1$ meV, close to the emission linewidth at low densities. The model reproduces the experimental narrowing of the exciton cloud with increasing density (Fig. 5a,b,e) and temperature (Fig. 5c,d,f), thus confirming the interpretation of the density-enhanced narrowing effect in terms of the screening of the trap disorder. At high densities, the trap behaves essentially ``disorder free''; compare experimental data with magenta line in Fig. 5e.

In conclusion, we have shown that a single diamond-shaped electrode creates a parabolic-like potential for indirect excitons. We observed collection of excitons to trap center with increasing density and  demonstrated that this effect is due to screening of the trap disorder by exciton-exciton interaction. At high densities, the trap behaves as a smooth parabolic potential which realizes a cold and dense exciton gas at the trap center.

This work is supported by ARO and NSF.

\end{document}